\documentclass[journal=jpcbfk,manuscript=article]{achemso}

\usepackage[version=3]{mhchem} 
\usepackage{epsfig}
\usepackage{graphics}
\usepackage{amsmath,amssymb,amsbsy}
\usepackage{mhchem}

\author{Matthew L. Strader}
\author{Hilton B. de Aguiar}
\author{Alex G. F. de Beer}
\author{Sylvie Roke}
\email{roke@mf.mpg.de}
\affiliation{Max-Planck-Institut fuer Metallforschung, Heisenbergstrasse 3, 70569 Stuttgart, Germany.}

\title[Spectroscopic Detection of Vesicle Bilayers in Water]
{Label-free Spectroscopic Detection of Vesicles in Water using Vibrational Sum Frequency Scattering}
\begin{document}
\begin{center}
\date{\today}
\end{center}
\begin{abstract}
Vibrational sum frequency scattering (SFS) has been used to study sub-micron, catanionic vesicles in solution.  The vesicles were synthesized from a binary mixture of dodecyltrimethylammonium bromide (DTAB) and sodium dodecylsulfate (SDS) surfactants in deuterated water, which spontaneously assemble into thermodynamically stable vesicles. The stability of these vesicles is attributed to a surfactant concentration asymmetry between the inner and outer bilayer leaflets. This concentration asymmetry should be observable by SFS due to local inversion symmetry-breaking. Signal corresponding to the symmetric sulfate stretch mode of the SDS head group is observed at 1044 cm$^{-1}$, indicating that there is indeed asymmetry in the local structure of the leaflets. The results indicate that it should be possible to measure the interfacial structure of liposomes in aqueous solution and study in-situ processes like the binding of sugars and proteins that are important for many processes in biophysical chemistry.\end{abstract}

\section{Introduction}

Cell membranes are astonishingly complex systems of lipids, proteins, carbohydrates and other components, which are often distributed asymmetrically between the exterior and interior bilayer leaflets of the cell membrane. The microscopic, structural details of the membrane influence a wide range of biological processes, such as ion transport, cell signaling and adhesion. The research of membrane structure has focused mostly on a number of cell membrane model systems, such as planar supported lipid bilayers (SLB) \cite{Castellana2006,Richter2006}, black lipid membranes and lipid monolayers \cite{Meier1986,Kaganer1999,Lalchev1999,Berkowitz2006}. The impermeable solid support required for SLB growth, by definition, limits the extent to which the membrane action of lipid bilayers can be studied. The study of free vesicles in solution, while immensely valuable, is often indirect, requiring a fluorophore to isolate membrane properties from the surrounding bulk media  \cite{Widengren1998,Schwille2001,Karukstis2003,Kahya2006,Silvius2006,Segota2006}.

Recently, it has been shown that Coherent Anti Stokes Raman (CARS) scattering can be used to identify the lipid density and the orientation of lipids (relative to the polarization of the laser beam) in large multilamellar vesicles \cite{Potma2005,Wurpel2005}. These measurements are label-free, but do not reflect leaflet-specific, chemically selective information. Sensitivity to leaflet asymmetry and charge were demonstrated with second harmonic scattering (SHS) \cite{Wang1996,Eisenthal2006}, a technique pioneered by the Eisenthal group. SHS was used to follow ionophore mediated transport of malachite green fluorophores through a liposome membrane \cite{Yan2000,Subir2008}.

Here, we show the possibility of detecting (asymmetry in) vesicle leaflets with chemical specificity, in-situ and label-free using vibrational sum frequency scattering. We obtain chemical information that is highly sensitive to the interfacial region that surrounds the vesicles, the asymmetry within the vesicle leaflets and the local surface charge around the vesicles.

Interfaces can be studied with vibrational Sum Frequency Generation (SFG), a technique in which visible (VIS) and infrared (IR) radiation mix to generate radiation at the sum of their frequencies \cite{GuyotSionnest1987,Harris1987}. Sum frequency (SF) radiation is only generated in media that lack inversion symmetry, because the simultaneous excitation of an IR transition and a Raman transition is required. SFG allows for the characterization of heterogeneous interfaces in a homogeneous bulk phase. The SFG process is greatly enhanced when the IR frequency is resonant with a molecular vibrational mode. Resonantly enhanced SFG also adds chemical specificity to the technique without the use of fluorescent labels \cite{Shen1989,Bain1995,Eisenthal1996,Bloembergen1999,Shultz2000,Richmond2001,Richmond2002,Chen2002,Shultz2002,Sioncke2003,Raschke2004,Simpson2004,Vidal2005,Lambert2005,Wang2005,Belkin2005,Kubota2007,Roke2009,Arnolds2010}. At the same time, the coherent nature of the method ensures a high sensitivity to molecular order, asymmetry, chirality, and surface charges. SFG has been employed successfully to investigate phospholipid monolayers \cite{Roke2003a,Bonn2004,Aroti2004,Ma2006,Ohe2007,Ma2007,Viswanath2009,Pavinatto2009,Casillas-Ituarte2010,Liljeblad2010} and recently also bilayers \cite{Liu2004,Liu2007a,Liu2008}.

In analogy to the method of second harmonic scattering \cite{Wang1996,Eisenthal2006,Schneider2007,Jen2009,Jen2010,Roke2009}, vibrational sum frequency scattering \cite{Roke2003,Roke2009} allows for the study of small particles in liquid solutions. SF scattering preserves the chemical sensitivity of SFG performed in reflection mode, while adding some additional specificity to surface charge and chirality \cite{Debeer2007,Roke2009,Debeer2011}. In an SF scattering experiment, an infrared (IR) and a visible (VIS) laser pulse are passed through a solution containing vesicles (see the inset of Fig. 1). The frequency of the IR radiation can be tuned to resonance with the vibrational modes of molecular groups in solution. These molecular groups can simultaneously undergo a resonant interaction with the IR field and a non-resonant Raman transition.  A resulting, second-order, sum frequency (SF) polarization is created in the medium, but due to selection rules, the polarization is limited to regions lacking inversion symmetry, e.g. surfaces. Coherent interference of SF radiation generated from this surface polarization will give rise to a scattering pattern in the far field. The scattering pattern depends on molecular structure and droplet size\cite{Roke2004,Dadap2009,Jen2009,Jen2010,Debeer2010,Debeer2011}.

To test whether vesicle membranes can be measured in-situ, we have made use of the versatility of catanionics. Catanionic vesicles are prepared from a mixture of cationic and anionic surfactants \cite{Kaler1989}. In contrast to kinetically formed lipid vesicles, catanionic surfactant mixtures assemble spontaneously into aggregates and are thermodynamically stable \cite{Segota2006,Lioi2009}. The favored aggregate can be either a micelle or a vesicle. The size of the resulting objects is controllable by moving within the catanion/anion/solvent ternary phase diagram. The vesicle is stabilized by disparate surfactant concentrations between the inner and outer bilayer leaflets, which results in a non-zero spontaneous bilayer curvature \cite{Safran1990,Safran1991,Seifert1993}. The leaflet concentration asymmetry has never been directly measured or confirmed, however. Accordingly, detection of SFS radiation from a catanionic vesicle solution would not only establish SFS as a viable technique for the study of vesicles and cell membranes, but also confirm bilayer asymmetry.

\section{Experimental Methods}
Catanionic mixtures composed of dodecyltrimethylammonium bromide (DTAB, Sigma-Aldrich >99\%) and fully deuterated sodium dodecylsulfate (SDS, Cambridge Isotope Laboratories > 98\% isotopic purity) were prepared in \ce{D2O} (Sigma-Aldrich > 98\% isotropic purity). Solution preparation was guided by the detailed phase diagram characterization by Herrington et al.\cite{Herrington1993}, in which surfactant concentrations are reported in weight percent.  A solution containing vesicles was prepared by mixing molar ratios of deuterated SDS, DTAB and \ce{D2O}, that correspond to the same molar ratios used for preparing a mixture of 0.35 wt\% DTAB, 0.65 wt\% non-deuterated SDS and 99 wt\% \ce{H2O}, as reported by Herrington et al.\cite{Herrington1993}. To facilitate comparison with the published phase diagram, our reported weight percentages have been scaled to remove deuteration effects. The solutions were characterized by dynamic light scattering with a Malvern ZS nanosizer.

The SF scattering measurements were performed by overlapping a 12 $\mu$J, broadband, femtosecond, IR pulse (see Ref \cite{Sugiharto2008a} for a description of the laser system), centered at 1050 cm$^{-1}$, with a 12-35 $\mu$J, picosecond, visible pulse centered at 803 nm in a  \ce{CaF2}/quartz sample cell (Hellma 106QS) with an optical path length of 100 $\mu$m that contained the vesicle solution. The IR and VIS pulses were focused down to a $\sim$0.4 mm beam waist under an angle of 15$^\circ$. The polarization of the IR beam is controlled by two BaF$_{2}$ wire grid polarizers. The polarization of the VIS beam was controlled by a polarizer cube and a half-wave plate. $p$-polarized beams are polarized parallel to the plane that holds the IR and VIS $\textbf{k}$-vectors (here the horizontal plane), whereas $s$-polarized beams are perpendicular to that direction (here, the vertical plane). Throughout the text polarization combinations are defined with a three letter code with the SF polarization first and the IR polarization last.

The SF scattered beam was collimated with a 0.5 inch diameter imaging lens (f=18 mm, Thorlabs LA1074B), and directed towards the detection system with two 2 inch silver mirrors. The imaging lens was placed at a scattering angle of 45$^{\circ}$, and the sample cell exit window was oriented perpendicular to the outgoing scattered light. The polarization of the SF signal was selected with a Glan-Taylor CaF$_2$ prism and spectrally filtered with two short-pass spectral filters (Thorlabs FES750 and Omega Optical 3RD-770) placed before the entrance slit of the spectrometer (Shamrock 303i, Andor Technologies). The SF signal was spectrally dispersed onto an intensified CCD camera (i-Star DH742, Andor Technologies), which employed a timing gate of 12 ns. The acquisition time of a single spectrum was 600 s. Recorded SF spectra are plotted as a function of IR wave number. The units on the y-axis of all graphs represents the counts of the I-CCD that has been baseline subtracted, and normalized by dividing the counts by the input energies of the IR and VIS pulses (in $\mu$J, measured right before the sample) and acquisition time (in s).

\section{Results and Discussion}
\begin{figure}[htpb]
    \epsfig{file=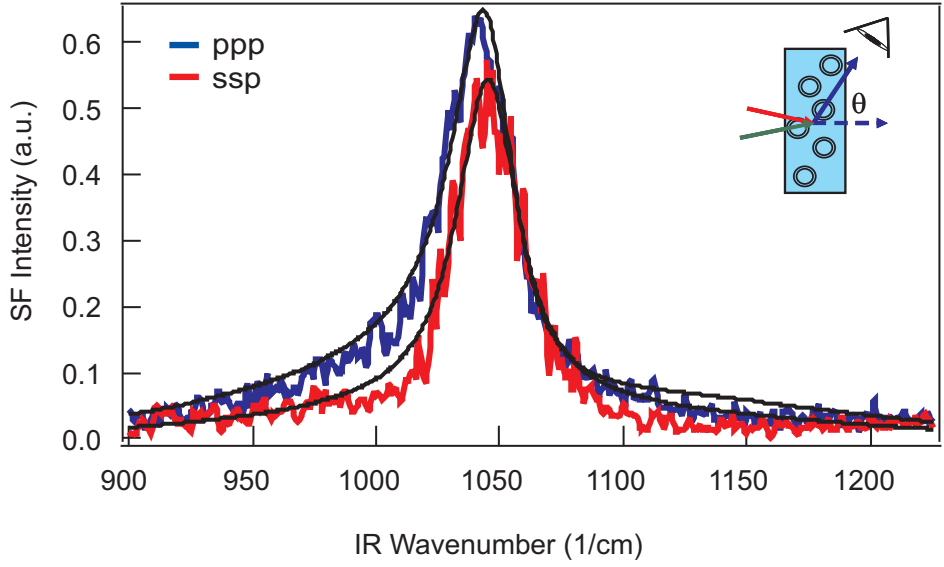,width=0.75\columnwidth}
    \caption{Vibrational sum frequency scattering intensity of the SDS symmetric sulfate stretch is plotted as a function of frequency for a 35 \% DTAB solution. This solution consists of vesicles with an average radius of 147 nm. The polarization combination of SF, VIS and IR radiation is specified by the legend. sps and pss signal could not be observed. Inset: illustration of SFS experiment: IR and VIS beams are transmitted through a cuvette. On the other side of the cuvette the scattered SF photons are collected.}\label{fig1}
\end{figure}
\ref{fig1} displays the SF scattering spectra of the vesicle solution. A single resonant feature is observed at 1044 cm$^{-1}$. This resonance is assigned to the symmetric stretch mode of the sulfate group of the SDS molecules. The spectra in the ssp and ppp polarizations were fit according to the following equation \cite{Roke2009}:
\begin{eqnarray}\label{sfsfit}
&I_{SFS}(\omega,\theta)\propto|\mathbf{E}_{IR}(\omega)(\sum_{n}\frac{{A}_{n}(\theta)}{(\omega-\omega_{0n})
+i\Upsilon_{0n}} + A_{NR}e^{i\Delta\phi})|^{2}\\
&A_{n}(\theta)=N_{s}F(\theta,\boldsymbol{\chi}^{(2)},R)\nonumber
\end{eqnarray}
which consists of a summation of all vibrational responses present in the spectral profile of the IR pulse. $n$ refers to a specific vibrational mode, with resonance frequency $\omega_{0n}$, and damping constant $\Upsilon_{0n}$. $E_{IR}$ is the envelope of the IR pulse, and $A_{NR}$ is the amplitude of the non-resonant contribution, which has a relative phase $\Delta\phi$ with respect to the vibrational resonances. $\theta$ is the scattering angle, defined as the angle between the sum of the incoming wave vectors and the detection direction. The amplitude of the scattered spectrum is determined by the molecular ordering and orientation (determined by the surface susceptibility $\boldsymbol{\chi}^{(2)}$) as well as the size (the radius, R) of the object. Ref. \cite{Debeer2010} gives a detailed explanation of how the scattered intensity depends on the molecular orientation.

The fit yields a resonance frequency of 1044 cm$^{-1}$ which had a Lorentzian half-width of $\Upsilon_{0n}$=16 cm$^{-1}$. The spectra also contain a broadly dispersive or non-resonant feature, which has different amplitudes for ssp and ppp polarization and might be tentatively assigned to the bending modes of the \ce{D2O} molecules at the interface. It is interesting to note that the observed symmetric stretch mode frequency lies $\sim$20 cm$^{-1}$ below that of earlier observed symmetric sulfate stretch modes: The symmetric \ce{SO3} stretch mode of the SDS molecule has been previously observed in reflection mode SFG measurements at 1070 cm$^{-1}$ from the planar air-water interface \cite{Johnson2005,Hore2005} and at 1080 cm$^{-1}$ at the hexadecane oil-in-water emulsion interface \cite{DeAguiar2010}. This difference demonstrates the chemical specificity of vibrational sum frequency scattering. The pronounced red shift compared to either of these single-amphiphile interfaces can be caused by e.g. a change in hydration state as has been suggested for lipid phosphate (\ce{PO2-}) stretching modes \cite{Ma2006}. Selective deuteration could also cause a shift of the vibrational mode.

To test the effect of deuteration on the sulfate stretch mode, we have recorded SF spectra of an emulsion prepared with 1 vol\% hexadecane in \ce{D2O} and stabilized with 1 mM SDS or 1 mM d-SDS. The spectral shape of the scattered spectrum was identical, except that in the case of d-SDS the vibrational mode was centered around 1051 cm$^{-1}$ instead of 1080 cm$^{-1}$. This indicates that the shift in frequency is not caused by a change in head group environment.

The presence of a sulfate signature suggests that the distribution of SDS in the inner and outer leaflet is not symmetric. We might therefore be able to measure as well the C-H modes of the alkyl chains of the DTAB in the vesicle. We have attempted to measure an SF spectrum in the spectral region of the C-H modes a number of times and were only able to distinguish a very weak spectrally broad signal once. Absence of a clear alkyl chain signal could point towards a highly disordered liquid-like structure of the alkyl chains inside the bilayer. Alternatively, SF signal can become very small if the orientation of the alkyl chains would be parallel with respect to the surface plane. Another possibility that would explain a combination of a relatively strong response from the S-O modes with a very low signal from the C-H modes is that we are detecting a quadrupole contribution (as described in the appendix of \cite{Debeer2009a}) from symmetrically distributed SDS molecules. Since the quadrupole contribution to the SF scattered signal is larger for molecular groups that are further apart it can be expected that the quadrupole signature is bigger for the S-O resonance than for the C-H resonances. Since a detectable quadrupole effect is unlikely for vibrational SFG experiments \cite{Held2002}, this explanation is not favored. It can, however, not be excluded.

To verify that the signal originates from an asymmetric vesicle bilayer and not from the micelles in the solution or surfactant adsorbed on the sample cell window, we have prepared a number of aggregate structures that contain identical total weight percentages of surfactant but correspond to different points in the cation/anion/solvent phase diagram. We have prepared samples with 0.325 (0.675) wt\%, 0.350 (0.650) wt\% and 0.375 (0.625) wt\% DTAB (SDS) and 99 wt\% \ce{D2O}. The average micelle and vesicle aggregate radii were determined by DLS to be 16 nm, 254 nm and 75 nm, respectively. These DLS measurements are consistent with a predominant aggregate structure of rod-like micelles (0.325 wt\% DTAB), multilamellar vesicles (0.350 wt\% DTAB), and unilamellar vesicles (0.375 wt\% DTAB) for the three solutions. Although these are the predominant species in the sample, it is likely that a vesicle solution also contains micelles \cite{Herrington1993}.
\begin{figure}[htpb]
    \epsfig{file=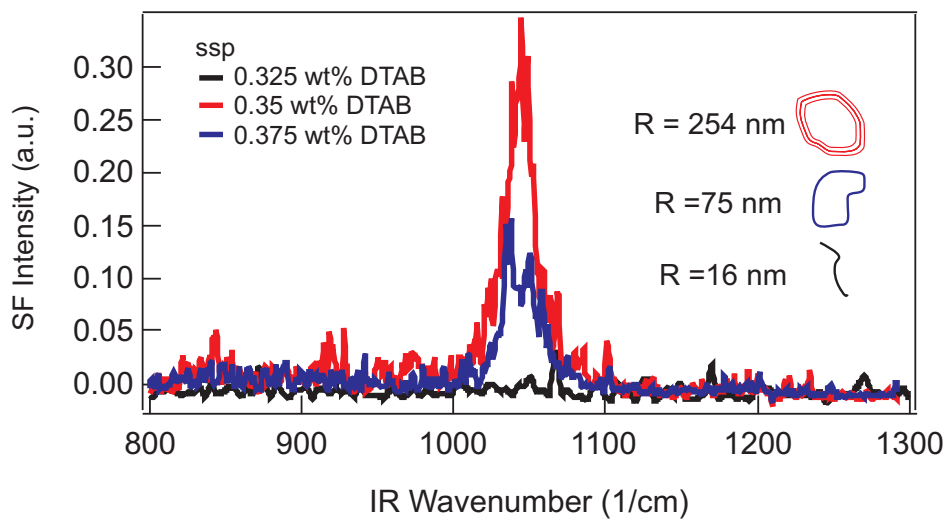,width=0.75\columnwidth}
    \caption{Vibrational sum frequency scattering intensity is plotted in the ssp polarization combination for various solutions with average aggregate radii specified in the legend. Spectral intensity has been normalized for VIS pulse power and acquisition time. The solution with 16 nm radius aggregates does not contain vesicles, and it does not produce detectable SF scattering radiation.}\label{fig2}
\end{figure}

\ref{fig2} displays the sum frequency scattering spectra in the ssp polarization combination of the three catanionic mixtures. It can be seen that the solution with small aggregates (rod-like micelles) does not produce measurable SF scattering intensity; in contrast, solutions with multilamellar and unilamellar solutions do produce a clear SF response. The SF signal corresponds to the signal of the sulfate stretch mode of the SDS. The resonant feature is therefore unambiguously assigned to SF scattering from vesicles, which directly confirms that the signal originates from the vesicle bilayer itself.

\section{Conclusions}

In summary, we have used the versatility of catanionic systems to unambiguously show that membranes of vesicles can be probed in water in-situ using vibrational sum frequency scattering with chemical specificity and a unique high sensitivity towards molecular order, asymmetry, chirality and surface charges. At the same time, our measurements have verified that the catanionic vesicle leaflet composition is asymmetric.

Catanionic vesicle systems are versatile and cheap. As such they are promising candidates for membrane mimicking systems, drug delivery, and electrokinetic separation (i.e. specific binding of polyelectrolytes such as DNA) \cite{Lioi2009}. Our approach will allow for a detailed understanding of the interfacial changes that occur and are required to specifically engineer selective binding. It may also provide a tool for studying and verifying solution theory that predicts the formation and stability of catanionic vesicles.

Our results also open up the road to study membrane processes in liposomes and more complex membrane systems with a high sensitivity to asymmetry, order, chirality and charge. As nonlinear optical spectroscopic methods are often complementary to linear spectroscopic techniques, linear light scattering, NMR \cite{Schiller2007}, and neutron scattering \cite{Prevost2009} it is likely that a wealth of new information may become available. Examples of processes that could benefit from such studies are: The working of ion pumps (while using simultaneously the ability of SFS to measure chemical structure and the exact surface potential), the asymmetric distribution of phospholipids across liposome bilayers, the accumulation of sugars on membranes and the transport of molecules across membranes, as well as cholesterol-induced changes in liposomes and membranes of small cells.

\begin{acknowledgement}
This work is part of the research programme of the Max-Planck Society. We thank the German Science Foundation (grant number 560398). MLS acknowledges the Alexander von Humboldt Foundation.
\end{acknowledgement}

\bibliography{./Strader_vesicles3.bbl}
\end{document}